\documentclass[floatfix,%
 reprint,
nofootinbib,
 amsmath,amssymb,
 aps,
floatfix,
]{revtex4-2}
\usepackage{caption}
\usepackage{graphicx}
\usepackage{siunitx}
\usepackage{braket}
\usepackage{subcaption}
\usepackage{makecell}
\usepackage{xcolor}
\usepackage{framed}
\usepackage{dcolumn}
\usepackage{bm}
\usepackage{float}
\usepackage{microtype}
\usepackage{mathrsfs}
\usepackage{float}
\usepackage[colorlinks,citecolor=blue,urlcolor=blue,linkcolor=blue]{hyperref}

\begin{document}

\preprint{APS/123-QED}

\title{Causal Structure of Spacetime Singularities and Their Observable Signatures}

\author
{Bina Patel$^{1}$}
\email{binapatel.maths@charusat.ac.in}
\author{Jahnvi Mistry$^{2}$}
\email{jahnvimistry25@gmail.com}
\author{Ayush Bidlan$^{3}$}
\email{i21ph018@phy.svnit.ac.in}
\author{Parth Bambhaniya$^{4}$}
\email{grcollapse@gmail.com}
\thanks{Corresponding author}

\affiliation{
 $^{1}$Department of Mathematical Sciences, P. D. Patel Institute of Applied Sciences, Charotar University of Science and  Technology (CHARUSAT), Changa, Gujarat 388421, India
}

\affiliation{
 $^{2}$Department of Physical Sciences, P. D. Patel Institute of Applied Sciences, Charotar University of Science and  Technology (CHARUSAT), Changa, Gujarat 388421, India
}
\affiliation{$^{3}$Department of Physics, Sardar Vallabhbhai National Institute of Technology, Surat 395007, Gujarat, India,}
\affiliation{$^{4}$Instituto de Astronomia, Geofísica e Ciências Atmosféricas, Universidade de São Paulo, IAG, Rua do Matão 1225, CEP: 05508-090 São Paulo - SP - Brazil}

\date{\today}
\begin{abstract}
We analyze the causal structure of horizonless compact objects via the light-cone geometry and conformal compactification of the Joshi-Malafarina-Narayan (JMN-1) and Janis-Newman-Winicour (JNW) spacetimes. Penrose diagrams reveal that JMN-1 undergoes a transition from timelike $(0<M_0<2/3)$ to null $(2/3<M_0<4/5)$ singularities, while JNW remains timelike throughout, in contrast to the spacelike singularity of the Schwarzschild spacetime. We show that photon spheres exist in Schwarzschild and JNW, but arise in JMN-1 only in the null singularity phase, establishing a direct link between causal character and null geodesic trapping. We further demonstrate that radial timelike geodesics develop turning points for certain parameter regimes in both JMN-1 and JNW spacetimes, indicating the emergence of effective repulsive behavior in the strong field region. These features lead to distinct strong field lensing and shadow signatures, potentially testable by very long baseline interferometric observations such as those of the Event Horizon Telescope.

\vspace{.5cm}

$\boldsymbol{Key words}$: Black Holes, Spacetime Singularities, Gravitational Collapse, Shadows, VLBI-EHT.
\end{abstract}

\maketitle
\section{Introduction}
The gravitational collapse of a spherically symmetric homogeneous dust cloud was first analyzed by Oppenheimer, Snyder, and Datt (OSD) in which the interior spacetime is described by a closed Friedmann geometry matched to an exterior Schwarzschild region \cite{Oppenheimer:1939ue,datt}. Within their framework, the collapse proceeds monotonically toward the formation of a curvature singularity. During the evolution, trapped two-surfaces form before the singular epoch is reached. The appearance of trapped surfaces directly implies that the future-directed null expansions become negative in the neighborhood of the center. As a result, the singularity lies entirely within the trapped region. It is, therefore, hidden from causal communication with the exterior spacetime region. More concretely, an event horizon forms, which acts as the causal boundary of this spacetime region.

In particular, the spacetime singularity is not an ordinary locus in spacetime, but rather it represents the boundary of the spacetime manifold \cite{Penrose1965}. All non-spacelike geodesics terminate there with a finite affine parameter $\lambda$. The associated Jacobi fields $\chi$ collapse, and the spacetime volume element along such congruences vanishes in the singular limit. The global causal structure of an asymptotically flat spacetime is conveniently described with respect to future null infinity $\mathcal{I}^{+}$. The event horizon is defined as the boundary of the causal past of $\mathcal{I}^{+}$, namely $\mathcal{H}^{+}=\partial J^{-}(\mathcal{I}^{+})$. It separates the spacetime region from which causal signals can reach distant observers from the region that remains permanently hidden. In collapse scenarios leading to black hole formation, the event horizon is a null hypersurface that encloses the trapped region and shields the high-curvature domain from external observation. Another important causal boundary in spacetime is the \textit{Cauchy horizon}, which is defined as the boundary of the domain of dependence of a spacelike hypersurface. For an initial spacelike hypersurface $\Sigma$, the domain of dependence $D(\Sigma)$ consists of all spacetime points whose physical evolution is uniquely determined by the initial data specified on $\Sigma$. The boundary of this region, known as the Cauchy horizon, marks the limit beyond which the classical evolution of the spacetime geometry cannot be uniquely predicted from the initial data. The presence of such a horizon therefore signals a breakdown of global determinism in general relativity (GR) \cite{Joshi:2024gog,Joshi:2011rlc}.

In GR, global predictability is ensured if the spacetime is globally hyperbolic, meaning that there exists a Cauchy surface whose domain of dependence covers the entire spacetime manifold. In particular, only a restricted class of highly symmetric solutions possess this property globally, including Minkowski spacetime, the Schwarzschild spacetime in its exterior region, and the Friedmann–Lemaître–Robertson–Walker (FLRW) cosmological models \cite{Joshi:2011rlc,Joshi:2024gog}. However, several physically important exact solutions of the Einstein equations admit inner Cauchy horizons. Notable examples include the charged black hole solution of Reissner–Nordström (RN) and the rotating black hole solutions of Kerr and Kerr–Newman. In their maximal analytic extensions, these spacetimes contain an inner horizon that acts as a Cauchy horizon separating regions where the deterministic evolution from initial data fails. Beyond this surface, the classical field equations no longer provide a unique continuation of the spacetime geometry from the initial data specified on a spacelike hypersurface. The occurrence of Cauchy horizons in such solutions indicates that global predictability is a strong condition in general relativity \cite{penrose1969} and need not be satisfied by all mathematically consistent solutions of the Einstein field equations \cite{joshi1993,Joshi:2011rlc,Joshi:2024gog}. This feature is closely related to fundamental questions concerning the validity of the Cosmic Censorship Conjecture (CCC) and the stability of Cauchy horizons under realistic perturbations.

The nature of a singularity may be characterized by the causal structure of the hypersurface on which incomplete causal geodesics terminate. A spacelike singularity possesses a spacelike boundary and lies entirely to the future of timelike observers, preventing any causal communication with the external spacetime. In contrast, timelike or null singularities have boundaries with corresponding causal character, allowing non-spacelike curves to originate from the singular region and propagate to other spacetime points. In particular, null singularities are generated by null geodesics and may permit outgoing radiation that reaches distant observers with arbitrarily large redshift \cite{Joshi:2020tlq,Dey:2020bgo}. These distinctions play a central role in determining the causal visibility of singularities and in classifying the possible end states of gravitational collapse.

The analysis of the OSD collapse motivated a broader conjecture proposed by Penrose, known as the Cosmic Censorship Conjecture (CCC) \cite{penrose1969}. The conjecture asserts that singularities arising in gravitational collapse are generically hidden within event horizons. This implies that the end state of the collapse is expected to be a black hole rather than a horizonless singularity. Despite sustained efforts, no precise and universally accepted mathematical formulation has been established. A general proof or disproof remains absent. The status of CCC remains a major open problem in classical and quantum gravity, with significant implications for black hole physics and relativistic astrophysics. In this context, the physical nature of ultra-compact objects remains unresolved.

A singularity is said to be \emph{naked} if future-directed non-spacelike geodesics originate from the singular region and reach future null infinity $\mathcal{I}^{+}$ \cite{joshi1993, Joshi:2011zm,Joshi:2011rlc}. In such cases, the singularity is not enclosed by an event horizon and may, in principle, influence the external spacetime. Extensive investigations of gravitational collapse have shown that the emergence of such horizonless singularities is not excluded within classical gravity and can arise from regular initial data under physically reasonable conditions \cite{Christodoulou:1986zr,Joshi:2011zm,Joshi:2011rlc,Joshi:2013dva,Joshi:2024gog,goswami2004,Deshingkar:1998ge,Bidlan:2026hee, Jhingan:2014gpa}. Moreover, the violation of CCC has also been discussed extensively in quantum gravity frameworks (see \cite{Hertog:2003zs, Bonanno:2016dyv,Bonanno:2017kta,Gutperle:2004jn} and references therein). The occurrence of horizonless spacetime singularities raises a fundamental question concerning the causal visibility and physical nature of these (ultra) compact objects. In particular, the mathematical classification of singularities according to their causal structure, i.e., spacelike, timelike, or null, may be associated with distinct physical and potentially observable signatures. Investigating this relation between causal structure and observable properties is therefore essential for determining whether astrophysical observations can distinguish black holes from horizonless compact objects containing different types of central singularities.

Recent advances in horizon-scale imaging through very long baseline interferometry have brought these questions into the domain of observational investigation \cite{Ayzenberg:2023hfw}. Observations by the Event Horizon Telescope (EHT) of supermassive compact objects, most notably Sgr~A* and M87*, have revealed bright ring-like emission structures that are commonly interpreted as the shadows of black holes \cite{EventHorizonTelescope:2019dse,EventHorizonTelescope:2022wkp,Vagnozzi:2022moj}. In particular, the EHT observations of Sgr~A* provide strong constraints on the geometry of the surrounding spacetime and are broadly consistent with the Kerr solution predicted by general relativity \citep{EHTC2022}. Nevertheless, current observations do not exclude all possible deviations from the Kerr black hole paradigm. Certain alternative compact object geometries can reproduce observational features similar to those expected from Kerr or Schwarzschild spacetimes. These considerations motivate continued theoretical investigations of compact object solutions, both within and beyond general relativity, especially in anticipation of future high-precision observations that may provide significant tests of the underlying spacetime geometry \cite{Vagnozzi:2022moj}.

Recent studies indicate that horizonless geometries, including certain classes of compact objects, can also produce shadow structures in high resolution images \cite{Saurabh:2023otl,gyulchev2019,Vagnozzi:2022moj,Joshi:2020tlq,Saurabh:2022jjv,Bambhaniya2021b,Bambhaniya2022,Bambhaniya:2025iqb,Bambhaniya:2021ugr,HassanPuttasiddappa:2025tji,Ortiz:2015rma}. We use the term ``shadow" to denote the apparent dark region in the observer's sky, irrespective of the presence of an event horizon. This observational degeneracy highlights the need for deeper theoretical investigations of alternative compact object models, focusing not only on image morphology but also on their causal structure, geodesic dynamics, pulsar timing delay, and energy extractions and tidal forces processes in the strong gravity regime \cite{Kalsariya:2024qyp,Madan2024,Bambhaniya:2026slo,Bambhaniya:2024hzb,Vertogradov:2024fva,Bambhaniya:2025qoe,Arora:2023ltv,Patel:2022jbk,Bambhaniya2019b}. Among the known horizonless compact object solutions, the JMN-1 \cite{Joshi:2011zm} and the JNW spacetimes \cite{Janis1968,Virbhadra:1997} provide particularly important probes of strong field gravity. Both arise as exact solutions of Einstein's equations with physically motivated matter sources and admit parameter ranges in which no event horizon forms \cite{Saurabh:2023otl,gyulchev2019}. Nevertheless, their geometrical and causal structures differ significantly. The JMN-1 solution represents the end state of gravitational collapse of anisotropic matter and can possess either null or timelike singularities depending on the compactness parameter $M_0$ \cite{Bambhaniya2023}. In contrast, the JNW geometry describes a static configuration sourced by a scalar field and contains a timelike singularity throughout its physically allowed parameter range. These spacetimes therefore provide a natural framework for investigating how observable properties of compact objects depend on the causal nature of the underlying singularity.

Particle dynamics in the vicinity of compact objects provides a complementary probe of strong gravitational fields \cite{Bambhaniya2019a,Bambhaniya2021a,Bambhaniya2024,Bambhaniya:2025xmu}. In extreme gravitational environments, particle collisions can achieve center of mass energies far exceeding those attainable in terrestrial accelerators \cite{Acharya:2023vlv,Patel:2023efv}. Such high-energy processes are believed to play a significant role in a variety of astrophysical phenomena, including relativistic jets, gamma-ray bursts, microquasars, active galactic nuclei, and high-energy cosmic rays \cite{GouveiaDalPino:2010phr,Singh:2014jma}. The efficiency of these acceleration mechanisms, as well as their potential observational signatures, depends sensitively on the underlying spacetime geometry. In particular, the presence of turning points, photon spheres, and regions of stable or unstable geodesic motion can strongly influence particle trajectories and collision outcomes \cite{Acharya:2023vlv,Patil:2010nt,Patil:2011aa,Patil:2011ya}. Investigating particle dynamics in horizonless spacetimes, therefore, provides an additional avenue for assessing the physical viability of alternative compact object models.

In this work, we present a systematic investigation of the JMN-1 and JNW spacetimes, focusing on three interconnected aspects: $(i)$ the causal structure of light-cones, $(ii)$ the conditions for shadow formation, and $(iii)$ particle acceleration in strong gravity. First, we analyze the global causal structure through the construction of light-cones and conformal compactification, allowing us to determine the causal nature of the singularities present in these geometries. Second, we examine null geodesics to identify the parameter regimes that give rise to shadow-like observational features. Third, we study timelike geodesics and particle collisions to evaluate the capability of these spacetimes to act as natural particle accelerators. Through this combined analysis, we aim to clarify which observable properties of compact objects are governed by the presence of horizons, which are controlled by photon spheres, and which arise from deeper aspects of spacetime geometry.

The analysis presented here indicates that shadow formation, the causal nature and strength of singularities, and particle acceleration or turning point constitute largely independent characteristics of compact object spacetimes and need not coincide within a single class of solutions. Therefore, observational signatures commonly associated with black holes may also arise in horizonless geometries. This highlights the importance of employing multiple observational probes when interpreting strong gravity phenomena. Distinguishing among these possibilities hence requires a synthesis of theoretical modeling and high precision observations. The present study contributes to this objective by providing a unified framework for analyzing strong gravity signatures in both black hole and horizonless compact object geometries.

This paper is organized as follows. In Sec.~\ref{sec2}, we analyze the causal structure of the JMN-1 spacetime through the construction of null cones and Penrose diagrams. In Sec.~\ref{sec3}, we perform a similar causal analysis for the JNW spacetime using null cones and conformal diagrams. In Sec.~\ref{sec4}, we investigate particle acceleration in the vicinity of the JMN-1 singularity. In Sec.~\ref{sec5}, we study particle acceleration along timelike geodesics in the JNW spacetime. In Sec.~\ref{sec6}, we examine the strength of the singularities present in the JMN-1 and JNW geometries. In Sec.~\ref{sec7}, we discuss the observational implications of these results in the context of horizon scale imaging with the EHT. Finally, in Sec.~\ref{sec8}, we summarize our results and present our conclusions. Throughout this paper, we adopt geometrize units in which $G=c=1$.

\section{Causal structure of the JMN-1 spacetime}
\label{sec2}
In the JMN-1 spacetime, the stress–energy tensor represents an anisotropic fluid with zero radial and non-zero tangential pressures. Nevertheless, the matter distribution satisfies all energy conditions, confirming its physical consistency. The corresponding interior JMN-1 metric is given by the following line element \cite{Joshi:2011zm},
\begin{equation}
    ds^{2}
    = - (1-M_{0})\left(\frac{r}{R_{b}}\right)^{\frac{M_{0}}{1-M_{0}}} dt^{2}
    + \frac{dr^{2}}{1-M_{0}}
    + r^{2} d\Omega^{2},
    \label{eq:JMN1_line element}
\end{equation}
where $d\Omega^2 = d\theta^2 + \sin^2 \theta\, d\phi^2$, and $R_b$ denotes the boundary radius at which the interior JMN-1 spacetime is smoothly matched to the exterior Schwarzschild geometry, thereby specifying the radial extent of the matter distribution.

\begin{figure}[t]
    \centering
    \includegraphics[width=\columnwidth]{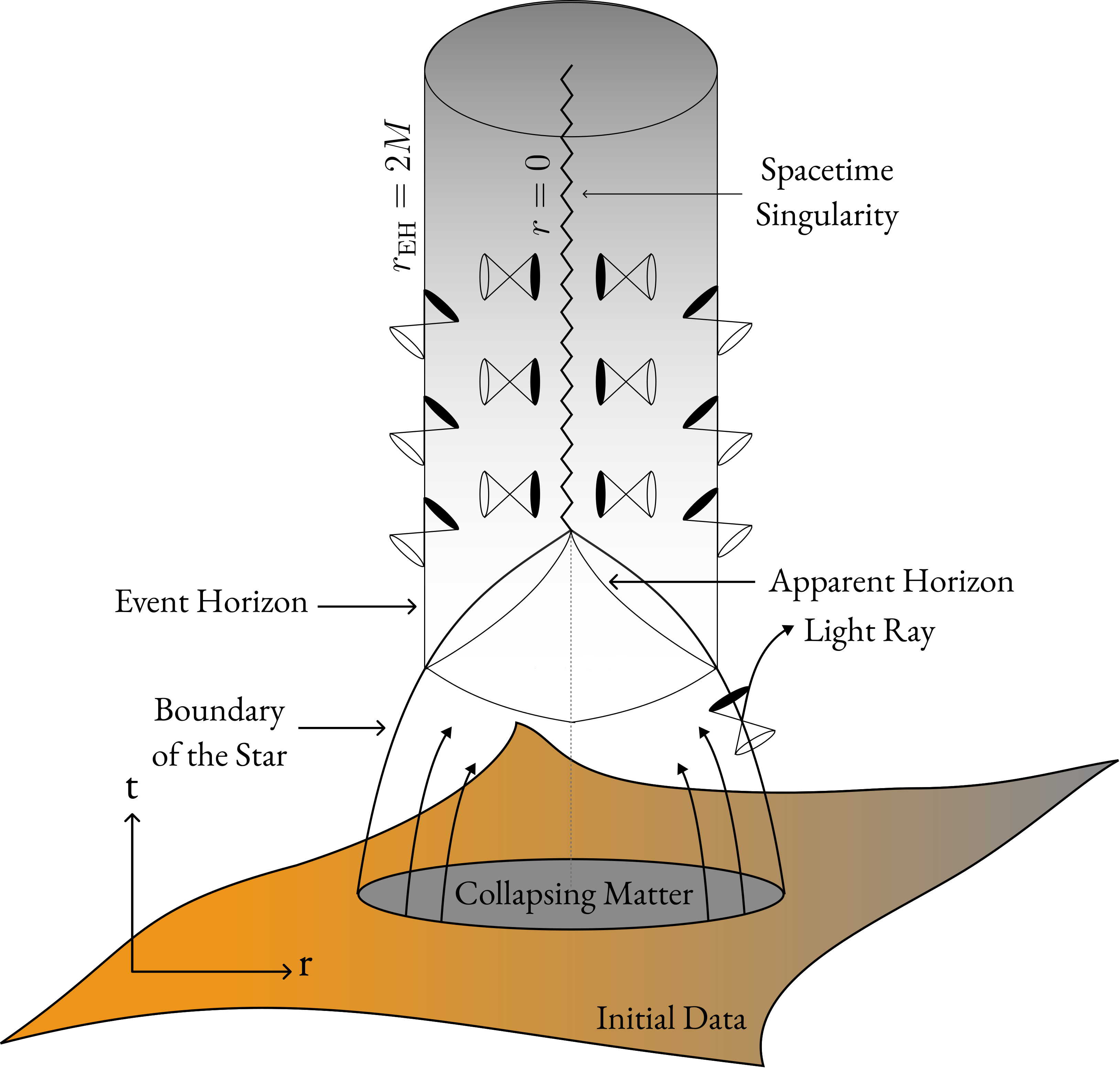}
    \caption{Schematic representation of the evolution of light-cone structures, highlighting the causal structure in the gravitational collapse of homogeneous dust.}
    \label{blackhole}
\end{figure}

\begin{figure}[t]
    \centering
    \includegraphics[width=\columnwidth]{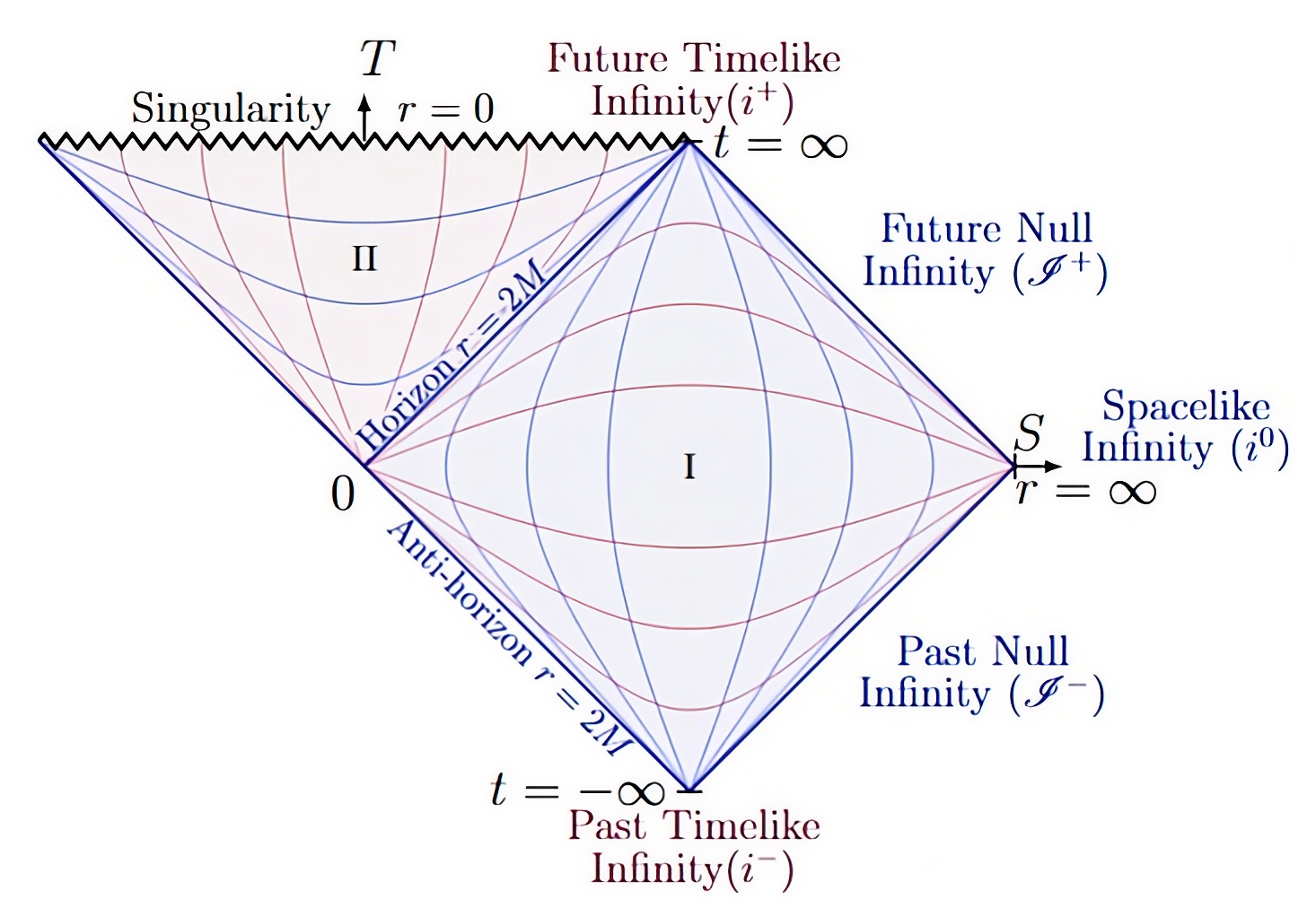}
    \caption{Penrose diagram of Schwarzschild black hole.}
    \label{fig:Sch_Penrose}
\end{figure}

We shows the schematic behavior of the light-cone and its associated causal structure within a gravitationally collapsing matter cloud leading to the formation of a Schwarzschild black hole in Fig. (\ref{blackhole}). The corresponding Penrose diagram of Schwarzschild spacetime, shown in Fig. (\ref{fig:Sch_Penrose}) for the completeness, uses conformal compactification to map the entire spacetime into a finite diagram while preserving the causal structure. Notably, the event horizon is located at $r=2M$ and the \textit{spacelike} spacetime singularity at $r=0$. 

The JMN-1 solution represents a spherically symmetric compact object of radius $R_b$, characterized by a dimensionless compactness parameter, and composed of gravitating matter with negligible non-gravitational interactions. Imposing the condition that the speed of sound remains subluminal leads to the constraint $R_b > 2.5 M$ \cite{Joshi:2011rlc}. The interior spacetime is smoothly matched at $r = R_b$ to the exterior vacuum Schwarzschild solution for $r \geq R_b$. Any static and spherically symmetric interior solution that is smoothly matched to a Schwarzschild geometry across a timelike hypersurface at $r = R_b$, must satisfy the appropriate junction conditions \cite{Bambhaniya2019a}. In particular, the continuity of the metric components and the extrinsic curvature across the boundary implies that the interior configuration is governed by the Tolman-Oppenheimer-Volkoff (TOV) equation \cite{Joshi:2011zm}, which characterizes the condition of hydrostatic equilibrium for a self gravitating fluid in general relativity,
\begin{equation}
    -\frac{dP_r}{dr}=(P_r+\rho )\frac{(4\pi P_r r^3+M)}{r(r-2M)}+\frac{2}{r}(P_r-P_{\theta}),
    \label{eq:TOV eq}
\end{equation}
where $P_{r}$ is radial pressure and $P_{\theta}$ is tangential pressure, and $\rho$ is the energy density. Within the JMN-1 interior spacetime, the matter distribution is intrinsically anisotropic, indicating that the pressure components along different spatial directions are not identical. In this study, we adopt the physically relevant assumption that the radial pressure vanishes, i.e., $P_{r}=0$, which implies that $\frac{dP_{r}}{dr}=0$. Incorporating this condition into Eq. (\ref{eq:TOV eq}) leads to a simplified equilibrium relation. The resulting TOV equation depends exclusively on the energy density $\rho$ and the tangential pressure components $P_{\theta}=P_{\phi}$. Hence, the hydrostatic equilibrium within the JMN-1 interior is maintained by the balance between gravitational forces governed by the density distribution and the supporting anisotropic tangential stresses.
\begin{equation}\label{eq:Ptheta}
    P_{\theta}=P_{\phi}=\frac{M_0 \rho}{4(1-M_0)}, \quad \rho=\frac{M_0}{R_b^2}.
\end{equation}
\begin{figure*}
    \centering
    \begin{subfigure}[!t]{0.45\linewidth}\vspace{1cm}
        \centering
        \includegraphics[width=\linewidth]{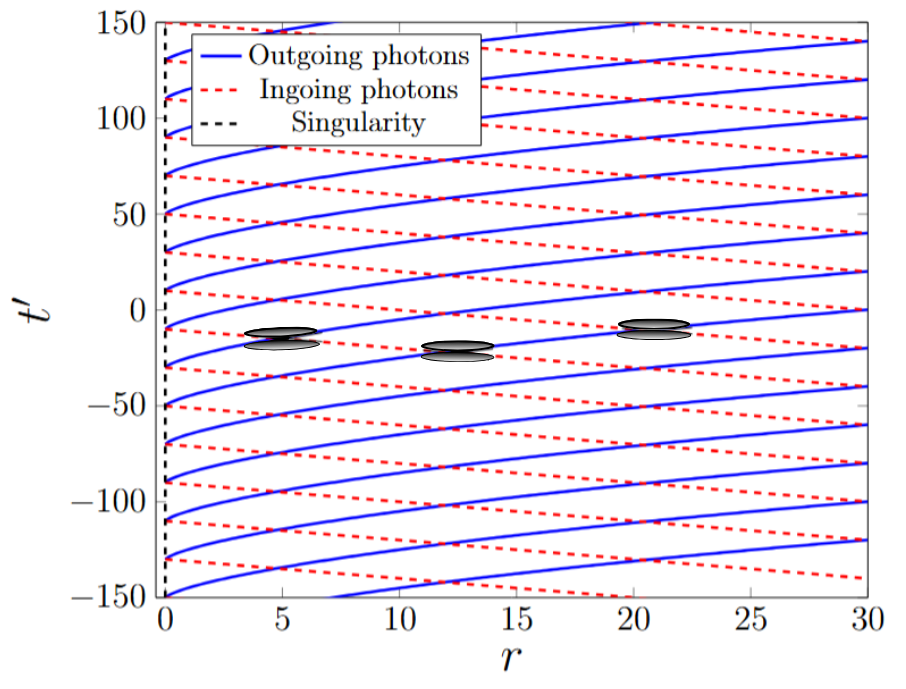} 
        \caption{Light-cone structure for $M_0=0.33$, $R_b=6.06$.} 
        \label{fig:Causal structure M0=0.33}
    \end{subfigure}
    \vspace{0.6cm}
    \begin{subfigure}[!t]{0.45\linewidth}
        \centering
        \includegraphics[width=\linewidth]{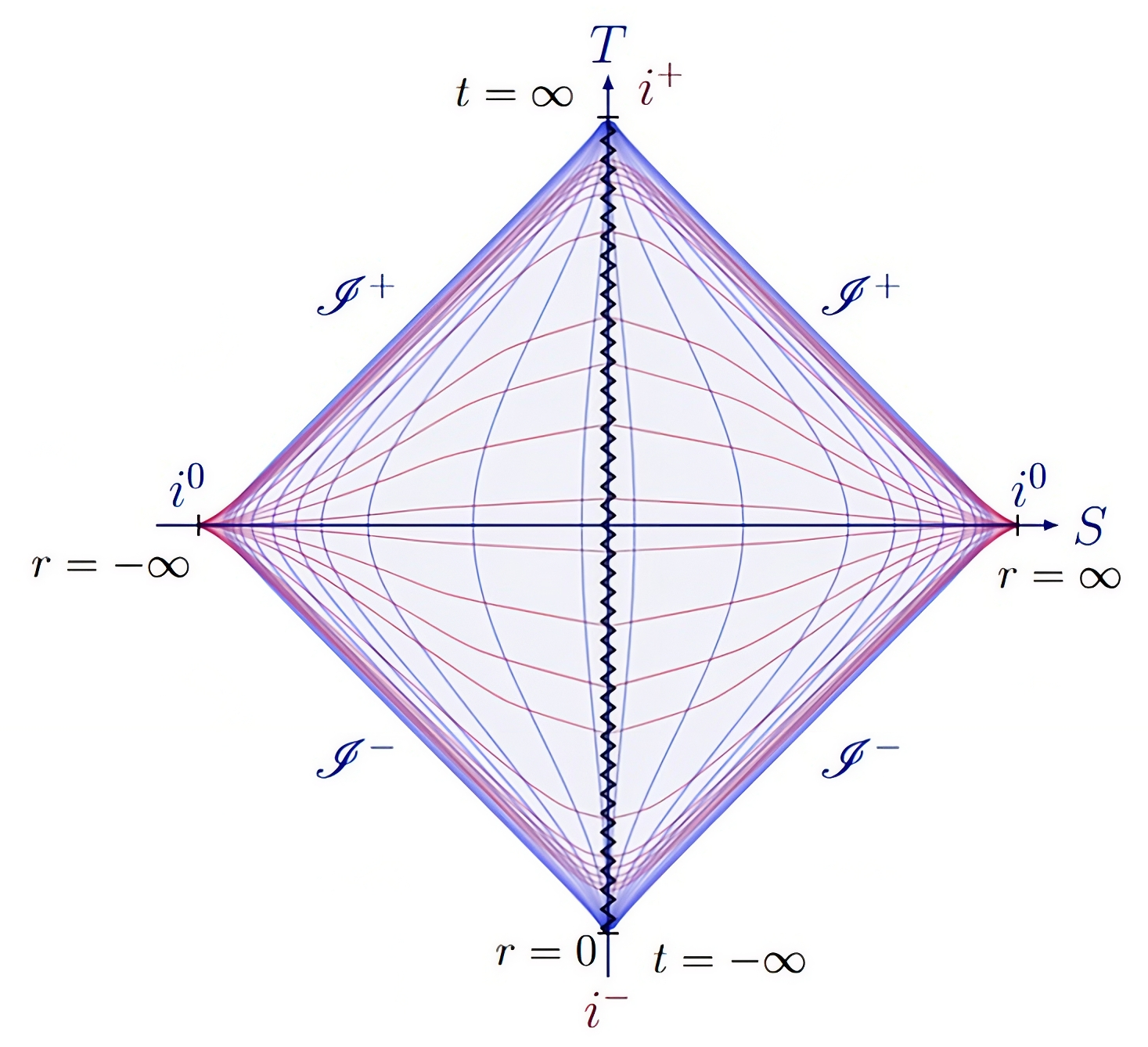}
        \caption{Penrose diagram for $M_0=0.33$, $R_b=6.06$.}
        \label{fig:Penrose_timelike}
    \end{subfigure}
        \begin{subfigure}[t]{0.45\linewidth}
        \centering
        \includegraphics[width=\linewidth]{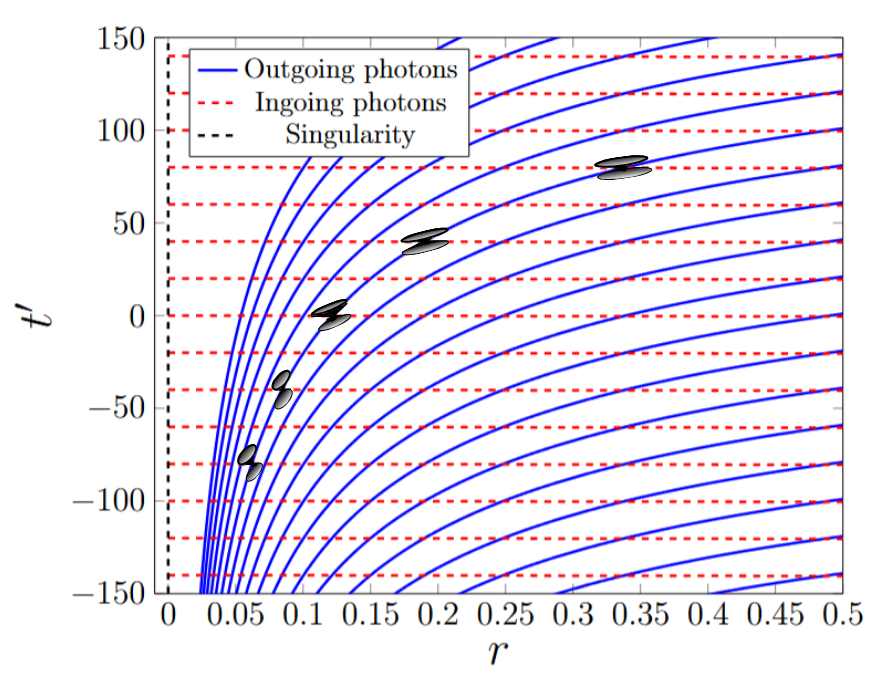}
        \caption{Light-cone structure for $M_0=0.75$, $R_b=2.67$.}
        \label{fig:Causal structure M0=0.75}
    \end{subfigure}
    \vspace{0.6cm}
    \begin{subfigure}[t]{0.45\linewidth}
        \centering
        \includegraphics[width=\linewidth]{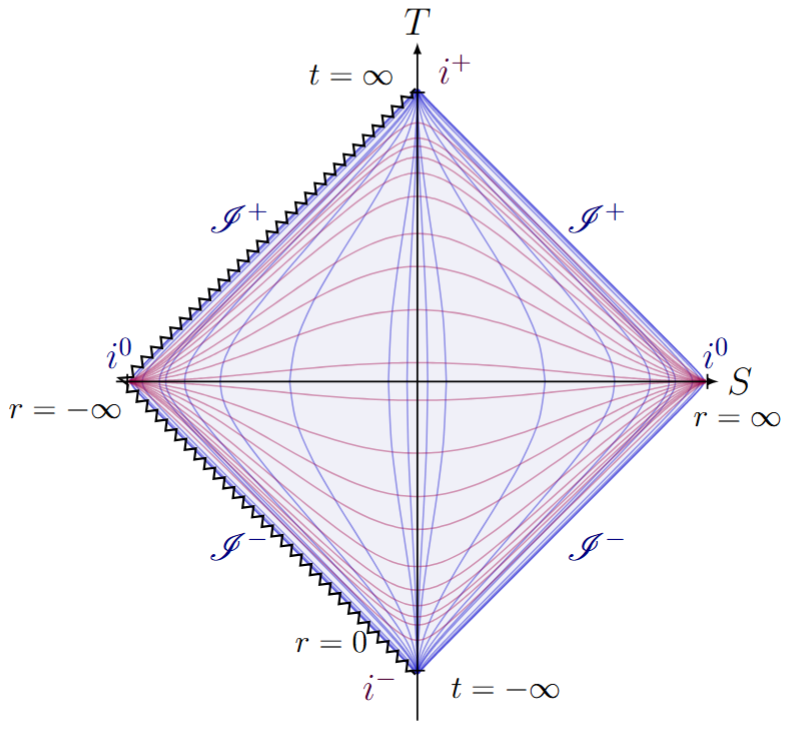}
        \caption{Penrose diagram for $M_0=0.75$, $R_b=2.67$.}
        \label{fig:Penrose_null}
    \end{subfigure}
    \caption{This figure represents the light-cone structures and corresponding Penrose diagrams of the JMN-1 spacetime, with the singularity being timelike for $M_{0} < \frac{2}{3}$ and null for $M_{0} > \frac{2}{3}$. Here, $r=0$ represents the spacetime singularity and the total mass is considered as $M=1$.}
    \label{Figure 3}
\end{figure*}

The causal structure of the JMN-1 spacetime can be analyzed through investigating the behavior of radial null geodesics near the singular region. 
This, in principle, can be done by imposing the condition $ds^{2}=0$, together with $d\Omega^2=0$, on the interior JMN-1 line element given in Eq.~\eqref{eq:JMN1_line element}. Under these conditions, the equation governing radial null trajectories in the $(t,r)$ plane reduces to the following form,
\begin{equation}
    \int dt=\pm \frac{1}{1-M_0}\int\left(\frac{r}{R_b}\right)^{-\frac{M_0}{2(1-M_0)}} dr.
    \label{eq: causal structrue eq}
\end{equation}
In the context of the light-cone structure, the positive sign represents outgoing null geodesics, while the negative sign corresponds to ingoing null geodesics. Integrating both sides of Eq.~(\ref{eq: causal structrue eq}) leads to the following expression,
\begin{equation}
    t=\pm \frac{2R_b}{2-3M_0}\left(\frac{r}{R_b}\right)^{\mathcal{A}}+ C,
    \label{eq:lightcone eq}
\end{equation}
where, $\mathcal{A}=\frac{2-3M_{0}}{2(1-M_{0})}$, and $C$ denotes the constant of integration. Here, both ingoing and outgoing null geodesics exhibit a non-trivial functional relationship between $t$ and  $r$, making the light-cone structure difficult to visualize in the $(t,r)$ coordinate. To resolve this difficulty, we adopt a construction analogous to the Eddington-Finkelstein (EF) coordinates, which provides a better description of the causal structure of the spacetime. Accordingly, for an ingoing radial null geodesic corresponding to the negative sign in Eq.~\eqref{eq:lightcone eq}, we define a new null coordinate $p$ as,
\begin{equation}
    p=t+r^*,
    \label{eq:adv-p-JMN-1}
\end{equation}
where, 
\begin{equation}
    r^{*}=\frac{2R_b}{2-3M_0}\left(\frac{r}{R_b}\right)^{\mathcal{A}}.
\end{equation}
Differentiating with respect to coordinate time, we obtain
\begin{equation}
    dt=dp-dr^*,
\end{equation}
and,
\begin{equation}
    dr^*=\frac{1}{1-M_0}\left(\frac{R_b}{r}\right)^{\frac{M_0}{2(1-M_0)}}dr,
\end{equation}
substituting these expressions into the JMN-1 metric given in Eq.~\eqref{eq:JMN1_line element} and simplifying, we obtain the line element
\begin{equation}
    ds^2=-f(r)dp^2+ g(r)\,dp \,dr+r^2d\Omega^2,
    \label{eq:AdvEF JMN-1 line element}
\end{equation}
where,
\begin{equation}
f(r)=(1-M_0)\left(\frac{r}{R_b}\right)^{\frac{M_0}{1-M_0}}, 
\quad
g(r)=2\left(\frac{r}{R_b}\right)^{\frac{M_0}{2(1-M_0)}},
\end{equation}
by setting $ds^2 = 0$ and $d\Omega^2 = 0$ in Eq.~\eqref{eq:AdvEF JMN-1 line element} obtain the radial null geodesics, which upon integration give
\begin{align}
    p&=C,
    \label{eq:p as continuous in advEF}\\
    p&=\frac{4R_b}{2-3M_0}\left(\frac{r}{R_b}\right)^{\mathcal{A}}+C.
    \label{eq:p as in advEF}
\end{align}
We introduce a new timelike coordinate $t'$ define as 
\begin{equation}
    t'=p-r,
\end{equation}
which gives
\begin{align}
      t'&=C-r,
    \label{eq:t'const}\\
    t'&=-r+\frac{4R_b}{2-3M_0}\left(\frac{r}{R_b}\right)^{\mathcal{A}}+C.
    \label{eq:t'outgoing}
\end{align}
Eq. \eqref{eq:t'const} shows that the ingoing null geodesics are regular, while the outgoing null geodesics remain well-behaved in the retarded EF coordinate. Since there is no event horizon in  the JMN-1 spacetime, a Kruskal-type extension is not needed. The advanced coordinate is given by Eq. \eqref{eq:adv-p-JMN-1}, and retarded coordinate is,
 \begin{equation}
     q=t- r^*.
     \label{eq:q as constant}
 \end{equation}
In this construction, the null coordinates $(p,q)$ are obtained from the original $(t,r)$ coordinates, where $p$ and $q$ respectively represent the outgoing and ingoing radial null geodesics, and both extend over an unbounded domain. To achieve spacetime compactification, bounded coordinates are introduced through the transformations,
\begin{equation}
    P=\tan^{-1}(p); \quad Q=\tan^{-1}(q),
\end{equation}
which map infinite ranges of the null coordinates to finite values while preserving the causal structure. The compactified coordinates $(P,Q)$ are then linearly transformed to define new coordinates $(T,S)$ as,
\begin{equation}
    T= \frac{1}{2}(P+Q); \quad S=\frac{1}{2}(P-Q),
\end{equation}
resulting in,
\begin{equation}
T=\frac{1}{2}\left(\tan^{-1}(p)+\tan^{-1}(q)\right),
\end{equation}
\begin{equation}
S=\frac{1}{2}\left(\tan^{-1}(p)-\tan^{-1}(q)\right).
\end{equation}

In specific terms, the parameter $M_0$ defines the compactness of the JMN-1 spacetime configurations and regulates the intensity of the overall effects caused by anisotropic pressure. When $M_0 < 2/3$, the spacetime does not admit a photon sphere. In this parameter regime, outgoing radial null geodesics emerging from the near by region of central singularity at $r \to 0$ can reach future null infinity, implying that the singularity is globally naked \cite{Shaikh:2018lcc}. The corresponding causal structure for a representative value $M_0 < 2/3$ is shown in Fig.~(\ref{fig:Causal structure M0=0.33}). 
The Penrose diagram in Fig.~(\ref{fig:Penrose_timelike}) confirms that the singularity is timelike. In the parameter range $2/3 \leq M_0 < 4/5$, a photon sphere forms and is located outside the matching radius $R_b$. The presence of the photon sphere substantially alters the behavior of null geodesics. Far from the singularity, i.e., $r\gg0$, the light-cone remains undistorted; however, as null rays propagate towards the singularity, the increasing spacetime curvature causes the light-cone to tilt inward, indicating a change in the local causal structure of the spacetime. Certain outgoing null trajectories become trapped near the photon sphere and are prevented from reaching future null infinity. Although the spacetime still contains a central horizonless singularity at $r \to 0^+$, the formation of the photon sphere restricts the causal propagation of null rays to distant timelike observers. The resulting causal structure for this parameter range is illustrated in Fig.~(\ref{fig:Causal structure M0=0.75}). The corresponding Penrose diagram is shown in Fig.~(\ref{fig:Penrose_null}), where the singularity appears as a \textit{null} (lightlike) boundary. However, no well defined Penrose diagram exists at $M_{0} = \frac{2}{3}$, as the system lies at a critical threshold marking the transition between distinct causal structures, separating regimes with qualitatively different null geodesic behavior.

\section{Causal Structure of the JNW Spacetime}
\label{sec3}
In this section, we examine the causal structure of the JNW spacetime. The JNW geometry is a static and spherically symmetric solution of Einstein's field equations sourced by a minimally coupled massless scalar field with vanishing scalar potential, $V(\Phi)=0$. This spacetime represents an extension of the Schwarzschild solution in the presence of a scalar field \cite{Janis1968}. The corresponding line element describing the JNW spacetime is given by \cite{Virbhadra:1997},
\begin{equation}
    ds^2
= -\left(1 - \frac{b}{r}\right)^{n}dt^2
+  \frac{dr^2}{\left(1 - \frac{b}{r}\right)^{n}}
+ r^2\left(1 - \frac{b}{r}\right)^{1-n} d\Omega^2,
\label{eq:JNWline element}
\end{equation}
where,
\begin{equation}
    b = 2\sqrt{M^2 + q^2}, \qquad
n = \frac{2M}{b},
\end{equation}
the corresponding scalar field $\Phi$ can be written as
\begin{equation}
    \Phi=\frac{q}{b\sqrt{4\pi}}\ln\left(1-\frac{b}{r}\right).
\end{equation}
In the limiting case $q \rightarrow 0$, which corresponds to $n=1$, the scalar field vanishes, i.e., $\Phi \rightarrow 0$. In this limit, the JNW spacetime reduces smoothly to the Schwarzschild geometry. When the scalar charge $q$ is nonzero, the spacetime departs from the Schwarzschild solution. The parameter $n = 2M/b$ decreases monotonically as the scalar charge increases. As a result, the scalar field modifies the spacetime geometry and the central curvature singularity at $r=b$, therefore, becomes horizonless. Larger values of the scalar charge lead to stronger deviations from the Schwarzschild spacetime \cite{Bambhaniya2019a}. 

The energy-momentum tensor of the scalar field satisfies the all energy condition everywhere in the spacetime \cite{Bambhaniya2019a,Bambhaniya2024}. However, for $0<n<1$, both the scalar field and the curvature invariants diverge at $r=b$, indicating the presence of a genuine curvature singularity. Since no event horizon forms to enclose this surface, the singularity remains globally visible. The limiting case $n=1$ corresponds to the Schwarzschild black hole, whereas the range $0<n<1$ describes a spacetime containing a spherical central singularity. Now, by setting $ds^2=0$ and $d\Omega^2=0$ in Eq. \eqref{eq:JNWline element}, the radial null geodesic satisfies,
\begin{equation}
dt=\pm \left(1 - \frac{b}{r}\right)^{-n} dr,
\label{eq: JNW Causal}
\end{equation}
where the positive and negative signs correspond to the outgoing and ingoing radial null geodesics, respectively. We can obtain time coordinate from Eq.~\eqref{eq: JNW Causal},
\begin{equation}
t=\pm \int \frac{dr}{\left(1 - \frac{b}{r}\right)^{n}}.
\label{eq: JNW_lightcone}
\end{equation}
In this context, both ingoing and outgoing null geodesics exhibit a complex relationship with r, making it difficult to visualize light-cone behavior. We employ null coordinates that remain constant along radial null geodesics, similar to the EF coordinates. For the radial ingoing null geodesic represented by a negative sign in Eq. \eqref{eq: JNW_lightcone}, we introduce a new coordinate denoted as $p$.

\begin{equation}
    p=t-r_*,
    \label{eq:JNW adv p}
\end{equation}
where, 
\begin{equation}
r_*=\int \frac{dr}{\left(1 - \frac{b}{r}\right)^{n}}.
\end{equation}
Taking the differential
\begin{equation}
    dt=dp-dr_*,
    \label{eq:JNW differential}
\end{equation}
where
\begin{equation}\label{qw}
    dr_*=\left(1 - \frac{b}{r}\right)^{-n}dr.
\end{equation}
By substituting above Eq. (\ref{qw}) into the JNW metric given in Eq. \eqref{eq:JNWline element} and simplifying, the line element becomes
\begin{equation}
    ds^2=-\left(1-\frac{b}{r}\right)^n dp^2+2dpdr+r^2d\Omega^2.
    \label{eq:AdvEF JNW line element}
\end{equation}
To understand the causal structure of this metric, we can examine the radial null geodesics within this new coordinate system. By substituting $ds^2 = 0$ and $d\Omega^2 = 0$ into Eq. \eqref{eq:AdvEF JNW line element}, we obtain two solutions, and integrating these solutions leads to
\begin{align}
    p&=C,
    \label{eq:JNW p as constant}\\
    p&=2\int\left(1-\frac{b}{r}\right)^{-n}dr+C.
    \label{eq:JNW p as in Adv}
\end{align}
Now using the null coordinate p as in Eq.\eqref{eq:JNW p as constant} and \eqref{eq:JNW p as in Adv}, we define a new timelike coordinate $t'$ through, $t'=p-r$, from which one obtains, 
\begin{align}
    t'&=C-r,
    \label{eq:JNW t' as constant}\\
    t'&=-r+2\int\left(1-\frac{b}{r}\right)^{-n}dr+C.
    \label{eq:JNW t' outgoing}
\end{align}
Here, Eq. \eqref{eq:JNW t' as constant} shows that all ingoing radial photons are continuous throughout the spacetime.
In the retarded EF coordinates, outgoing null geodesics remain continuous. As the JNW spacetime does not contain an event horizon in this case, a Kruskal extension is not required. The corresponding advanced EF coordinates are given as in Eq. \eqref{eq:JNW adv p}, and retarded EF coordinates are given by
\begin{equation}
    q=t-r_*.
\end{equation}
Since the null coordinates $(p,q)$ are unbounded, a compactification is performed by introducing the bounded coordinates,
\begin{equation}
    P=\tan^{-1}(p); \quad Q=\tan^{-1}(q),
\end{equation}
 which map the spacetime to a finite domain. The resulting compactified coordinates $(P,Q)$ are then subjected to a rotation to define a new set of coordinates $(T,S)$, given as
 \begin{equation}
 T=\frac{1}{2}(P+Q);\quad   S=\frac{1}{2}(P-Q), 
 \end{equation}
resulting in,
\begin{equation}
T=\frac{1}{2}\left(\tan^{-1}(p)+\tan^{-1}(q)\right).
\end{equation}
\begin{equation}
S=\frac{1}{2}\left(\tan^{-1}(p)-\tan^{-1}(q)\right).
\end{equation}

\begin{figure*}
    \centering
    \begin{subfigure}[b]{0.45\linewidth}
        \centering
        \includegraphics[width=\linewidth]{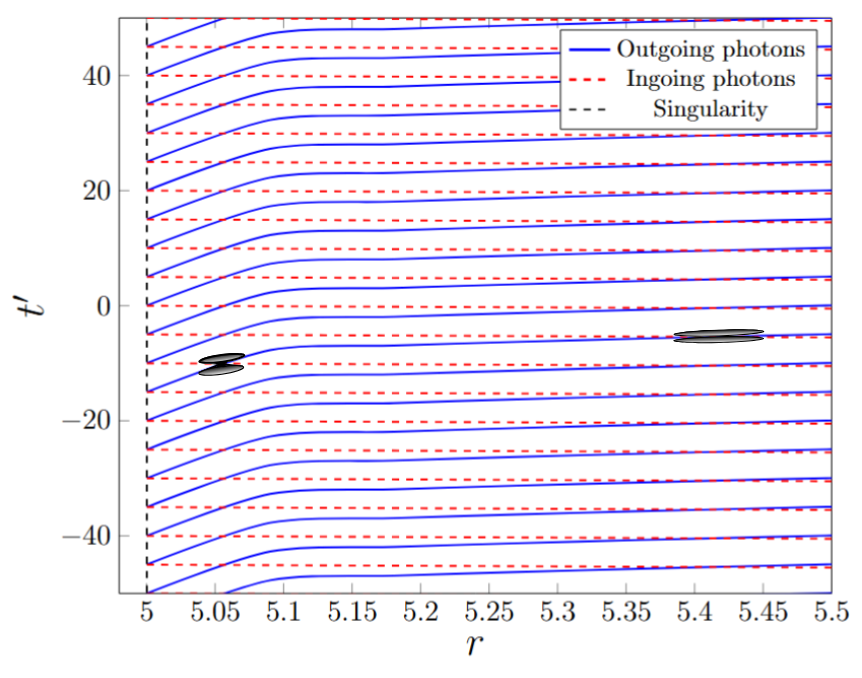}
        \caption{Light-cone structure for $n=0.4$, $b=5$.}
        \label{fig:Causal structure n=0.4}
    \end{subfigure}
    \begin{subfigure}[b]{0.45\linewidth}
        \centering
        \includegraphics[width=\linewidth]{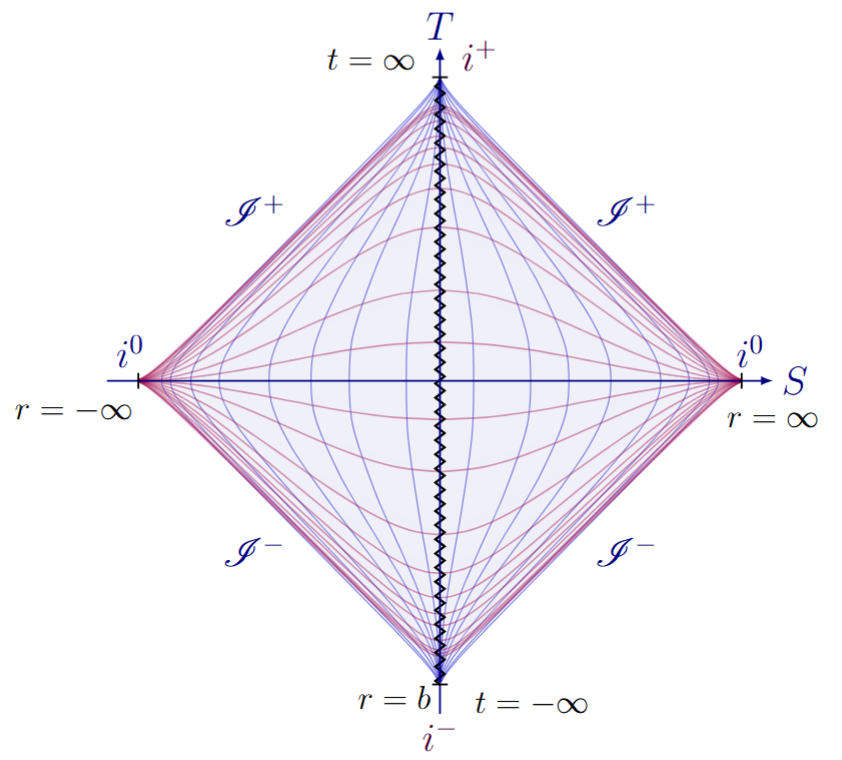}
        \caption{Penrose diagram for $n=0.4$, $b=5$.}
        \label{fig:JNW penrose-n=0.4}
    \end{subfigure}
    \begin{subfigure}[b]{0.45\linewidth}
        \centering
        \includegraphics[width=\linewidth]{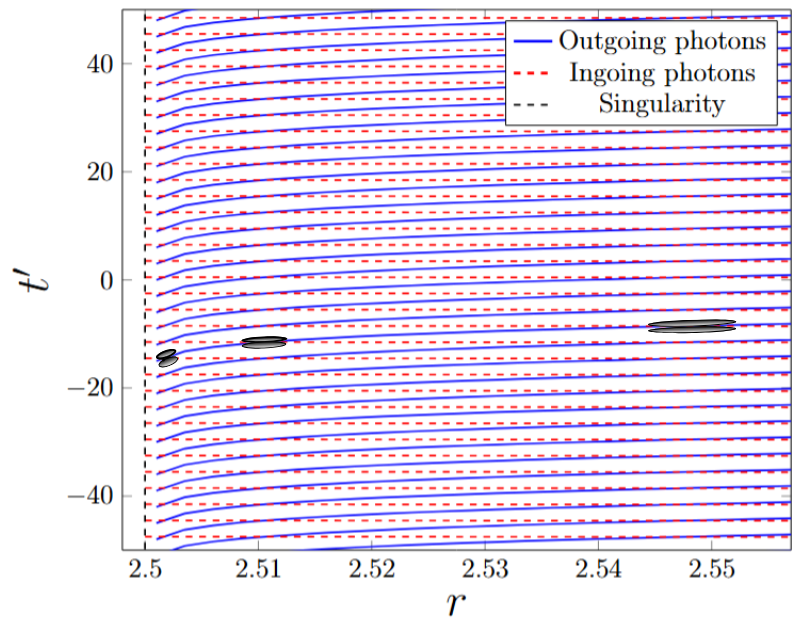}
        \caption{Light-cone structure for $n=0.8$, $b=2.5$.}
        \label{fig:Causal structure n=0.8}
    \end{subfigure}
    \begin{subfigure}[b]{0.45\linewidth}
        \centering
        \includegraphics[width=\linewidth]{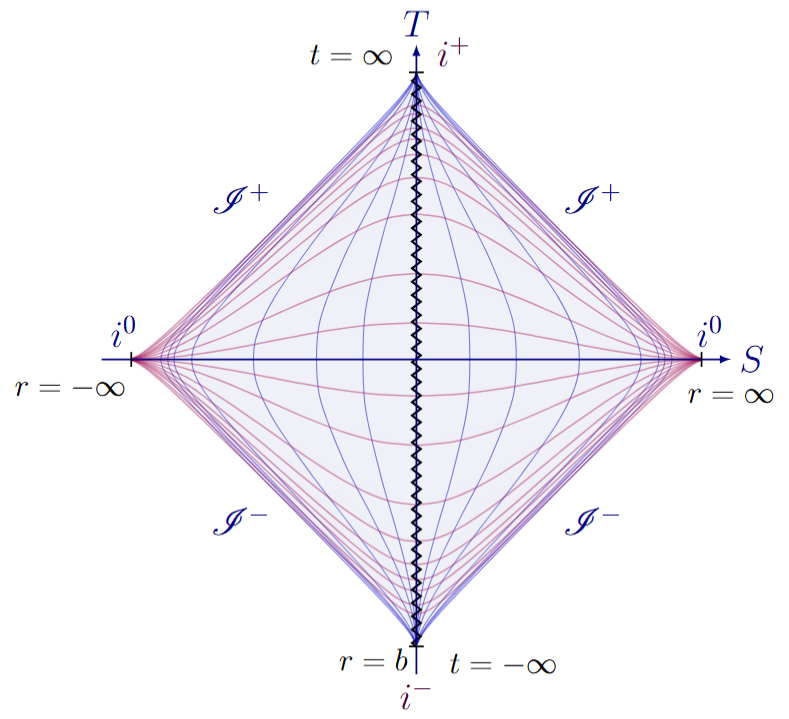}
        \caption{Penrose diagram for $n=0.8$, $b=2.5$.}
        \label{fig:JNW penrose-n=0.8}
    \end{subfigure}
    \caption{This figure shows the light-cone structures and corresponding Penrose diagrams of the JNW spacetime, with the singularity being timelike for $0 < n < 1$, where $r=b$ represents the spherical singularity. The total mass is considered as $M=1$.}
    \label{Figure 4}
\end{figure*}
 
The causal structure of the JNW spacetime can be understood by analyzing the behavior of radial null geodesics described by Eq. \eqref{eq:JNW t' as constant} and \eqref{eq:JNW t' outgoing}. As a result, the singular surface located at $r=b$ remains horizonless. Near the singular surface $r=b$, the light-cones become strongly distorted but do not close to form a trapped region. Therefore, radial null geodesics starting arbitrarily close to the singularity are able to propagate outward and reach null infinity without any obstruction. As illustrated in Fig.~(\ref{fig:Causal structure n=0.4}), for $n=0.4$ as one of the example of $0<n<0.5$ range, null geodesics emerging arbitrarily close to the singular surface and extend smoothly toward null infinity. The absence of any trapped region or horizon in this case confirms that no event horizon is formed, and therefore the singularity is globally naked.

In the regime $0.5<n<1$, the dynamical influence of the scalar field becomes comparatively weaker relative to the cases with smaller values of $n$. Despite this reduction in strength, the scalar field continues to play a significant role in determining the local causal structure of the spacetime. In particular, near the central singularity its contribution to the spacetime curvature remains sufficiently strong to substantially distort the light-cone structure. Thus, the trajectories of outgoing null geodesics experience a pronounced deflection in this region, reflecting the strong bending of light-cones induced by the scalar field. This behavior indicates that, even in this parameter range, the scalar field continues to influence the propagation of null rays and the accessibility of the singularity within the spacetime. This behavior modifies the propagation of radial null geodesics, causing them to deviate from their otherwise regular trajectories. As shown in Fig.~(\ref{fig:Causal structure n=0.8}) for $n=0.8$ as one of the example of $0.5<n<1$ range, the null geodesics show bending as they pass close to the singular surface at $r=b$ while still propagating outward toward null infinity. The corresponding Penrose diagram of the JNW spacetime for different values of $n$ is shown in Fig.~(\ref{fig:JNW penrose-n=0.4}) and ~(\ref{fig:JNW penrose-n=0.8}). 

The conformal compactification shows that the curvature singularity at $r=b$ appears \textit{timelike} in nature in the Penrose diagram for the parameter range $n\in(0,1)$. This indicates that the singularity remains causally connected to the external universe, allowing ingoing causal curves to terminate at the singular surface while outgoing null geodesics can emerge from regions arbitrarily close to it. In JMN-1 collapse model, photon trapping regions can change the causal nature of the singularity. However, no such transition occurs in the JNW spacetime. Although the causal structure shows some quantitative differences for different values of $n$, such as variations in the tilting of light-cones and the behavior of geodesics near the singularity, the qualitative nature of the singularity itself remains the same. In particular, the absence of an event horizon throughout the parameter range $n\in(0,1)$ ensures that the singularity is globally naked. This persistent timelike nature indicates that the non-zero scalar field alters the spacetime curvature but does not lead to the formation of a horizon, and therefore the spherical singular surface remains causally connected to the exterior spacetime.

\section{Turning Points of Timelike Geodesics Near the JMN-1 Singularity}
\label{sec4}
The line element describing the JMN-1 spacetime is given in Eq.(\ref{eq:JMN1_line element}). The metric functions are positive and regular for all $r>0$, provided the parameter lies in the range $0<M_0<1$. Radial null trajectories are determined by the condition $ds^2=0$. Restricting to purely radial motion, this condition reduces to
\begin{equation}
-f(r)\,dt^2 + g\,dr^2 = 0 ,
\end{equation}
where $g=1/(1-M_0)$. Solving this relation yields the coordinate speed of outgoing radial null rays,
\begin{equation}
\frac{dr}{dt} = (1-M_0)\left(\frac{r}{R_b}\right)^{\alpha},
\qquad
\alpha \equiv \frac{M_0}{2(1-M_0)} .
\label{eq:drdt_null}
\end{equation}
This explicit relation plays an important role in determining the causal structure of the spacetime. The prefactor $(1-M_0)$ is strictly positive for all admissible values $0<M_0<1$, while the power-law factor $(r/R_b)^{\alpha}$ remains positive for every $r>0$. Thus, the coordinate velocity $dr/dt$ is positive throughout the interior region. This immediately implies that the JMN-1 spacetime does not contain an event horizon within the interior, since there exists no hypersurface at which outgoing radial null geodesics satisfy $dr/dt=0$. However, the radial dependence of the factor $(r/R_b)^{\alpha}$ significantly affects the orientation of the local light-cone structure. As $r \rightarrow 0$, the power-law term decreases rapidly, causing the coordinate speed of outgoing null rays to become progressively smaller. Geometrically, this corresponds to a strong inward tilting of the light-cones near the central singularity, indicating that the outward propagation of null geodesics becomes increasingly suppressed in this region.

To understand this behavior in more detail, consider the factor $(r/R_b)^{\alpha}$ as the radius decreases. For any fixed compactness parameter $M_0\in(0,1)$, the exponent $\alpha$ is positive. Therefore, $(r/R_b)^{\alpha}$ decreases monotonically as $r$ decreases. As a result, the coordinate speed of outgoing light, $dr/dt$, also decreases toward smaller radii. When $r$ becomes much smaller than $R_b$, the power-law factor $(r/R_b)^{\alpha}$ becomes very small. Consequently, $dr/dt$ becomes correspondingly small. Physically, this means that outgoing null directions near the central region are almost vertical in a $t-r$ diagram. Future directed light rays then have only a very small outward radial velocity. The light-cones therefore tilt increasingly inward as $r$ decreases toward the singularity. This behavior represents a continuous modification of the causal structure rather than the formation of a horizon. Outgoing rays still satisfy $dr/dt>0$ for all $r>0$, but the rate at which their radius increases becomes arbitrarily small sufficiently close to the singularity.

The strength of this inward tilting depends on the compactness parameter $M_0$ through the exponent $\alpha$. When $M_0$ is small, $\alpha$ is also small and the factor $(r/R_b)^{\alpha}$ remains relatively large over most of the interior. Outgoing rays therefore retain a significant outward speed until they approach very close to $r=0$. As $M_0$ increases, the exponent $\alpha$ becomes larger and the factor $(r/R_b)^{\alpha}$ decreases more rapidly with decreasing $r$. The value $M_0=2/3$ plays a special role. For $M_0>2/3$, the exponent $\alpha$ exceeds unity. In this regime, the suppression of $dr/dt$ with decreasing $r$ becomes very strong even at moderate radii. The light-cones then begin to tilt inward at much larger radii compared to smaller values of the compactness parameter. This extended inward tilting leads to an effective trapping of photons and is closely related to the shadow features associated with the spacetime.

To connect the interior causal behavior with observable shadows, the exterior photon sphere must also be considered. In the exterior Schwarzschild region, the photon sphere is located at $r=3M$. Photons that reach this radius with insufficient outward radial momentum cannot escape to infinity. Instead, they are either captured or remain near the unstable circular orbit for a long time before eventually escaping or falling inward. When the JMN-1 interior is sufficiently compact, outgoing null rays emerging from the interior may not possess enough outward radial momentum to cross the photon sphere. Such rays therefore do not contribute to the radiation observed at large distances and are effectively removed from the observable flux. 

For $M_0>2/3$, the suppression of $dr/dt$ inside the JMN-1 region becomes very strong. Null rays that reach even moderate interior radii emerge with extremely small outward velocity. These rays cannot overcome the Schwarzschild photon sphere and are either captured or redirected inward. As a result, they do not reach distant observers. In this way, the shadow arises from the combined effect of the interior causal structure and the exterior photon sphere. The interior geometry strongly reduces the outward propagation speed of photons, while the photon sphere prevents the weakly outgoing rays from escaping to infinity.

The motion of massive particles and the possibility of having turning points are determined by the corresponding effective potential. For equatorial motion, the conserved energy per unit mass $e$ and angular momentum per unit mass $L$ determine the radial equation in the form
\begin{equation}
    \left(\frac{dr}{d\tau}\right)^2 + V_{\mathrm{eff}}(r)=E,
\end{equation}
where, $E=\frac{e^2-1}{2}$. The effective potential in the JMN-1 interior is given by
\begin{equation}
    V_{\mathrm{eff}}(r)=\frac{1}{2}\left[(1-M_0)\left(\frac{r}{R_b}\right)^{\frac{M_0}{1-M_0}}
\left(1+\frac{L^2}{r^2}\right)-1\right].
\end{equation}
Turning points occur when $(dr/d\tau)^2=0$. Consider particles that fall in from rest at infinity. For such particles, the energy satisfies $E=0$. The turning-point condition then gives a relation between the radius $r$ and the angular momentum $L$,
\begin{equation}
L(r)=\pm r\left[\frac{R_b^{\frac{M_0}{1-M_0}}}{(1-M_0)\,r^{\frac{M_0}{1-M_0}}}-1\right]^{1/2}.
\label{eq:L_of_r}
\end{equation}
This expression is real only if the quantity inside the square brackets is positive. Therefore, turning points exist only at radii where
\begin{equation}
    \frac{R_b^{\frac{M_0}{1-M_0}}}{(1-M_0)\,r^{\frac{M_0}{1-M_0}}} > 1 .
\end{equation}
Since $R_b$ and $M_0$ are fixed parameters of the model, this condition imposes a nontrivial constraint on the allowed values of $r$ and on the compactness parameter $M_0$.

\begin{figure}[h]
    \centering
    \includegraphics[width=1.0\linewidth]{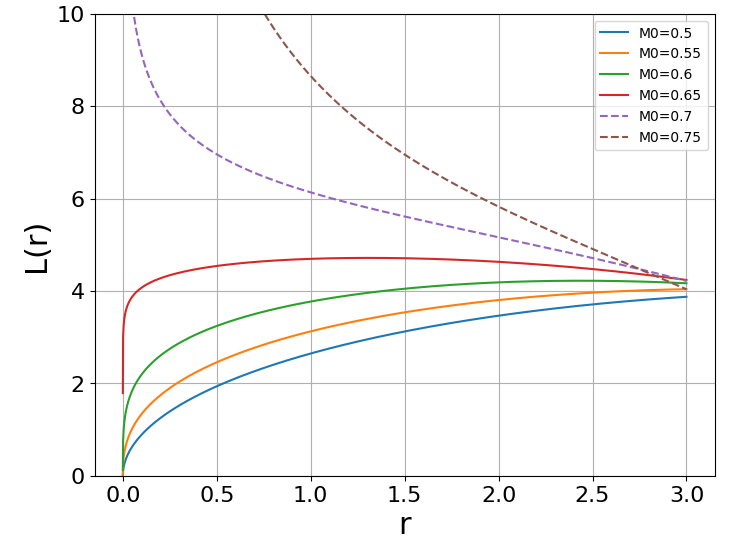}
    \caption{Angular momentum $L(r)$ required for radial turning points as a function of the radial coordinate $r$ for different values of the parameter $M_0$. The color bar indicates the corresponding values of $M_0$.}
    \label{fig:JMN-1 angular momentum vs radius}
\end{figure}
A detailed analysis of Eq.~\eqref{eq:L_of_r} shows that the function $L(r)$ can possess a minimum. This minimum corresponds to the critical angular momentum required for a particle to turn its radial motion at some radius inside the matter region. The angular momentum $L(r)$ required for radial turning points in the JMN-1 spacetime for $\frac{1}{2}<M_0<\frac{4}{5}$ is shown in Fig.~(\ref{fig:JMN-1 angular momentum vs radius}). Such a minimum exists within the interval $0<r\le R_b$ only when the compactness parameter lies in the range $1/2 \le M_0 \le 2/3$. If $M_0<1/2$, the radius at which the minimum occurs lies outside the boundary radius $R_b$. In this case, any turning point would occur outside the interior region. Therefore, such trajectories cannot produce turning points within the JMN-1 matter region and cannot lead to collisions inside the interior. If $M_0>2/3$, the quantity inside the square root in Eq.~\eqref{eq:L_of_r} never becomes positive for $r\le R_b$. As a result, $L(r)$ is not real in this region and no turning point exists inside the interior. Particles with the required angular momentum to turn back their motion therefore do not exist. All timelike geodesics that enter the interior region plunge monotonically toward the central singularity. This result is consistent with the analysis presented in \cite{Acharya:2023vlv}.

Combining the null and timelike analyses leads to the following physical picture. For compactness in the range $1/2 \le M_0 \le 2/3$, the interior region supports turning point geodesics. A particle arriving from infinity with a suitable angular momentum can decelerate and reverse its radial motion inside the matter region. This produces an outgoing timelike trajectory that can intersect with an ingoing particle. Such encounters allow high center of mass energy collisions to occur in the strong field region. As $M_0$ increases within this interval, the turning point radius shifts toward smaller values of $r$. At the same time, the spacetime curvature becomes stronger near the interaction region. Both effects tend to enhance the maximum achievable collision energy. For $M_0>2/3$, the situation changes qualitatively. The interior causal structure becomes strongly tilted inward. Outgoing null directions are then strongly suppressed. In this regime, no timelike turning points exist inside the matter region. Particles that enter the interior therefore continue to move inward and plunge toward the singularity. Hence, the outgoing trajectories required for head-on, high-energy collisions cannot form as shown in \cite{Acharya:2023vlv,Patil:2010nt,Patil:2011aa,Patil:2011ya}.

In particular, a visible shadow can form for $M_0>2/3$. In this regime, the interior geometry strongly suppresses the outward propagation of photons. At the same time, the exterior photon sphere captures rays with weak outward momentum. Therefore, photons that originate in the interior fail to reach distant observers and contribute to the shadow appearance. However, the same causal suppression also affects massive particles. The strong inward tilting of the light-cones prevents the formation of outgoing timelike trajectories. Therefore, turning point geodesics do not exist inside the matter region when $M_0>2/3$. Without such trajectories, particles cannot reverse their radial motion within the interior. Thus, the ingoing and outgoing particle paths required for high-energy collisions cannot occur. This implies that high center of mass energy collisions are suppressed within the shadow region and hence no shock waves or photosphere like structure will appear, that is, within the region bounded by the photon sphere ($r \leq 3M$). This result is consistent with the analysis presented in \cite{Broderick:2024vjp,Acharya:2023vlv}. The turning point radii for representative values of the compactness parameter $M_0$ are listed in Table~\ref{tab:turning_points}. These values correspond to particles with the minimum angular momentum required to reverse their radial motion inside the JMN-1 interior. The radii are obtained from Eq. (35) by substituting $L=L_{\rm min}(r)$ together with the matching condition $M_0 R_b = 2M$. As the compactness parameter increases within the allowed interval, the turning point shifts to smaller radii, indicating that the reversal of particle trajectories occurs deeper in the strong field region. In agreement with the analytical constraints discussed above, such turning points exist only in the range $1/2 \le M_0 \le 2/3$.
\begin{table}[h]
\centering
\small
\renewcommand{\arraystretch}{1.4}
\setlength{\tabcolsep}{18pt}
\caption{Turning point radii ($r_{\rm turn}$) within the $R_b$ in the JMN-1 spacetime for particles with $L_{\rm min}$.}
\label{tab:turning_points}
\begin{tabular}{|c|c|}
\hline\hline
\textbf{$M_{0}$} & $r_{\rm turn}$ \\
\hline
0.50 & $4.00\,M$ \\
\hline
0.55 & $3.23\,M$ \\
\hline
0.60 & $2.43\,M$ \\
\hline
0.65 & $1.31\,M$ \\
\hline
\end{tabular}
\end{table}

\section{Turning Points of Timelike Geodesics Near the JNW Singularity}
\label{sec5}
The line element of the JNW spacetime is given in Eq. (\ref{eq:JNWline element}). The dynamics of timelike geodesics are governed by the corresponding effective potential. This potential also determines the conditions under which a particle can accelerate and develop a turning point. Because the spacetime is stationary and spherically symmetric, the motion of a test particle admits two conserved quantities, the energy per unit mass $e$ and the angular momentum per unit mass $L$. By restricting the motion to the equatorial plane ($\theta = \pi/2$), the effective potential in the JNW spacetime can be written as
\begin{equation}
    V_{\mathrm{eff}}(r)=\frac{1}{2}\left[\left(1-\frac{b}{r}\right)^{n}
\left(\frac{L^2}{r^2\left(1-\frac{b}{r}\right)^{1-n}}+1\right)-1\right].
\end{equation}
The turning point condition then provides a relation between the radius $r$ and the angular momentum $L$,
\begin{equation}
L(r)=\pm r\left[\left(1-\frac{b}{r}\right)^{1-2n}-\left(1-\frac{b}{r}\right)^{1-n}\right]^{1/2}.
\label{eq:L_of_r-JNW}
\end{equation}
\begin{figure}[htbp]
    \centering
    \includegraphics[width=1\linewidth]{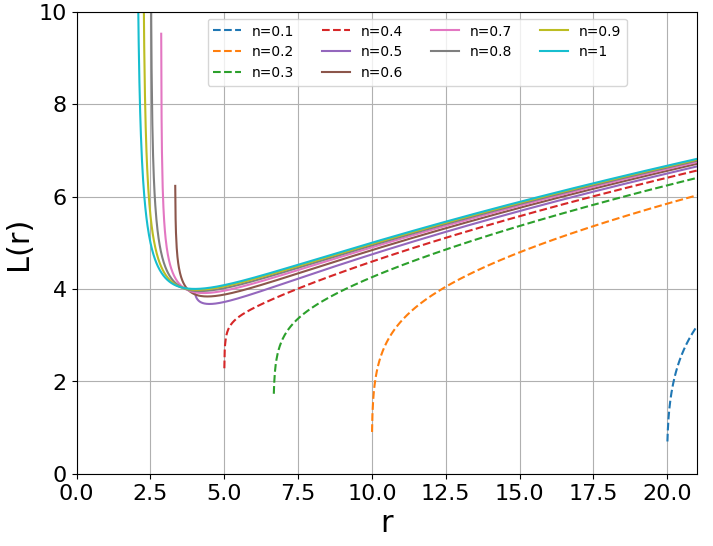}
    \caption{Angular momentum $L(r)$ required for radial turning points as a function of the radial coordinate $r$ for different values of the parameter $n$. The color bar indicates the corresponding values of $n$.}
    \label{fig:JNW angular momentum vs radius}
\end{figure}

The angular momentum $L(r)$ required for radial turning points in the JNW spacetime for $0<n<1$ is shown in Fig. \ref{fig:JNW angular momentum vs radius}. Real turning points appear only when $n \geq 0.5$. For smaller values of $n$, the expression for $L(r)$ becomes imaginary (nonphysical) in the region $r>b$. The case $n=0.5$ represents a critical configuration in which the angular momentum remains finite at the singularity. In this situation, $L(r)$ attains a minimum at a finite radius. Larger values of $n$ shift this $L(r)$ minimum toward smaller radii and increase its magnitude. Such behavior allows particles arriving from infinity to reverse their radial motion closer to the singularity and participate in high-energy collisions. This method and result are consistent with the analyses presented in \cite{Patil:2011aa,Patil:2010nt}.

In the case of $n = 0.5$, the angular momentum remains a finite value at the singularity $r = b$. The function $L(r)$ reaches its minimum value $L_{\min}=3.674$ at $r=4.5M$. Symmetry of the equations under the transformation $L\rightarrow -L$ implies identical extrema for positive and negative values of $L$. Particles arriving from infinity with angular momentum in the interval $-L_{\min}<L<L_{\min}$ can therefore penetrate deep into the strong gravitational region near $r\sim b$ before turning back their radial motion. Particles with larger angular momentum are instead scattered at larger radii. On the other side, values $n>0.5$ cause the location of the minimum of $L(r)$ to move progressively closer to the singularity as $n$ increases. The corresponding magnitude of $L_{\min}$ also grows. This behavior reflects the strengthening of the gravitational field near the singularity. Particles arriving from infinity with suitable angular momentum can decelerate, reach a finite turning point radius, and collide head-on with other radially infalling particles. Such interactions can generate extremely large center of mass energies in the strong-field region, analogous to the Ba\~{n}ados--Silk--West (BSW) mechanism \cite{Patil:2010nt,Patil:2011aa,Patil:2011ya}. These high-energy collisions may give rise to shock waves or an effective photosphere, leading to enhanced electromagnetic emission \cite{Broderick:2024vjp}. As a result, this may, in principle, lead to observable signatures in EHT shadow imaging, such as excess luminosity or bright substructures near the photon sphere, while the overall shadow remains largely determined by the underlying photon orbit geometry.

Turning point radii for representative values of the parameter $n$ are listed in Table \ref{tab:turning_points_jnw}. These radii correspond to particles possessing the minimum angular momentum required to reverse their radial motion in the JNW spacetime. The values are obtained from Eq. \eqref{eq:L_of_r-JNW} by substituting $L=L_{\min}(r)$. Increasing $n$ shifts the turning point radius gradually inward toward $4M$, indicating that particle trajectory turning occurs closer to the central region. This result is in agreement with the analytical condition derived above, according to which such turning points exist only for $n \geq 0.5$.
\begin{table}[h]
\centering
\small
\renewcommand{\arraystretch}{1.4}
\setlength{\tabcolsep}{18pt}
\caption{Turning point radii ($r_{\rm turn}$) in the JNW spacetime for particles with $L_{\rm min}$.}
\label{tab:turning_points_jnw}
\begin{tabular}{|c|c|}
\hline\hline
\textbf{$n$} & $r_{\rm turn}$ \\
\hline
0.5 & $4.50\,M$ \\
\hline
0.6 & $4.43\,M$ \\
\hline
0.7 & $4.30\,M$ \\
\hline
0.8 & $4.18\,M$ \\
\hline
0.9 & $4.08\,M$ \\
\hline
1.0 & $4.00\,M$ \\
\hline
\end{tabular}
\end{table}

\section{Tipler's Strength of JMN-1 and JNW Singularities}
\label{sec6}
A spacetime singularity is characterized by the existence of at least one incomplete causal geodesic. However, in the context of collapsing matter, a stronger physical requirement is often imposed, namely that any extended object falling into the singularity is crushed to \textit{zero} volume. A singularity satisfying this condition is said to be gravitationally strong, in the sense introduced by Tipler \cite{Tipler:1977zza}.
Let $(\mathcal{N}, g)$ be a smooth spacetime manifold and let $\Gamma(\lambda)$ be a causal geodesic defined on an interval $[\lambda_0, 0)$, where $\lambda$ is an affine parameter, and the endpoint $\lambda = 0$ corresponds to the singularity. Let $\chi_i$ denote a set of linearly independent Jacobi vector fields along $\Gamma$, orthogonal to the tangent vector of the geodesic. The wedge product of these Jacobi fields defines a volume element 
\begin{equation}
    \mathcal{V}= \chi_1 \wedge \chi_2 \wedge \chi_3.
\end{equation}
The singularity is said to be strong in the sense of Tipler if this volume element vanishes in the limit $\lambda \to 0$. Clarke and Kr\'{o}lak provided a sufficient condition for the occurrence of a Tipler strong curvature singularity in terms of the growth of the curvature along causal geodesics \cite{CLARKE1985127}.  In particular, at least along one null geodesic with affine parameter $\lambda$ (such that $\lambda\rightarrow0$ as the singularity is approached), the following condition must hold,
\begin{equation}\label{Tipler}
\lim_{\lambda\rightarrow0}\lambda^{2}R_{\mu\nu}K^{\mu}K^{\nu}>0.
\end{equation}
Here, $R_{\mu\nu}$ is the Ricci curvature tensor, $K^{\mu}=dx^{\mu}/d\lambda$ are the tangent vectors to the null geodesics, and $x^{\mu}$ is the spacetime coordinate. This condition puts a lower bound on the growth of the Ricci curvature tensor along the null geodesic. Within our analysis, we will consider the interior spacetime of the singularity to be described by the JMN-1 spacetime geometry with the line element given by Eq. (\ref{eq:JMN1_line element}). Using the null geodesic condition, $g_{\mu\nu}K^{\mu}K^{\nu}=0$, we obtain the following relationship between $K^{0}$ and $K^{1}$ components of the null tangent vector,
\begin{equation}\label{Strn}
    K^{0}=\pm\frac{1}{(1-M_{0})}\left(\frac{R_{b}}{r}\right)^{\frac{M_{0}}{2(1-M_{0})}}K^{1}.
\end{equation}

In order to examine the strength of the singularity, we require the Ricci tensor components $R_{00}$ and $R_{11}$ for the JMN-1 metric, which are given by,
\begin{equation}
    R_{00}=-\frac{M_{0}(M_{0}-2)}{4r^{2}}\left(\frac{r}{R_{b}}\right)^{\frac{M_{0}}{1-M_{0}}}, 
\end{equation}
\begin{equation}
   R_{11}=\frac{(2-3M_{0})M_{0}}{4r^{2}(M_{0}-1)^{2}}. 
\end{equation}
The Tipler strong singularity criterion requires the left hand side of the inequality to approach a positive finite value as given in Eq.(\ref{Tipler}), substituting the relevant components, we obtain
\begin{equation}
\lim_{\lambda\rightarrow0}\lambda^{2}\left[\frac{M_{0}}{r^{2}(1-M_{0})}\right]\left(\frac{dr}{d\lambda}\right)^{2}.
\end{equation}
To determine whether this quantity takes a positive finite value near the singular limit $\lambda\rightarrow0$, we examine the behavior of 
\begin{equation}
\lim_{\lambda\rightarrow0}\left(\frac{dr}{d\lambda}\right).
\end{equation}
This is obtained by writing the Lagrangian for the radial null geodesics as 
\begin{equation}
\begin{split}
\mathcal{L}^{\text{null}}_{\text{JMN-1}}\equiv\frac{1}{2}\Bigg[-(1-M_{0})\left(\frac{r}{R_{b}}\right)^{\frac{M_{0}}{1-M_{0}}}\left(\frac{dt}{d\lambda}\right)^{2}\\+\frac{1}{(1-M_{0})}\left(\frac{dr}{d\lambda}\right)^{2}\Bigg]=0.
\end{split}
\end{equation}
The equation of motion for $r$ is obtained from the Euler–Lagrange equation corresponding to $\mathcal{L}^{\text{null}}_{\text{JMN-1}}$. This gives,
\begin{equation}
    \frac{d^{2}r}{d\lambda^{2}}+\frac{M_{0}}{2r(1-M_{0})}\left(\frac{dr}{d\lambda}\right)^{2}=0.
\end{equation}
We consider an ansatz solution to this nonlinear differential equation, with a radial coordinate $r(\lambda)\sim\tilde{\alpha}(\lambda-\beta)^{\gamma}$ valid for all values of the parameter $M_{0}$. Here $\tilde{\alpha}$, $\beta$ and $\gamma$ are positive constants. The first derivative then takes the form, 
\begin{equation}
    \frac{dr}{d\lambda}\sim(\tilde{\alpha}\gamma)(\lambda-\beta)^{\gamma-1}.
\end{equation}
For simplicity, we set the constant $\beta=0$. Substituting these expressions into the Tipler criterion, we obtain,
\begin{equation}
\lim_{\lambda\rightarrow0}\lambda^{2}R_{\mu\nu}K^{\mu}K^{\nu}=\frac{M_{0}}{(1-M_{0})}>0.
\end{equation}
Since $0 < M_{0} < 1$, the above limit is finite and positive; therefore, the central singularity satisfies the Tipler strong curvature condition. This result indicates that the JMN-1 spacetime exhibits a strong singularity for all values of the compactness parameter in the allowed range $M_{0}\in(0,1)$.

We now consider the JNW spacetime, whose line element is given by Eq. (\ref{eq:JNWline element}).
The relation between the components $K^{0}$ and $K^{1}$ of the null tangent vector for the JNW metric is given by,
\begin{equation}
    K^{0}=\pm\left(1-\frac{b}{r}\right)^{-n}K^{1}.
\end{equation}
The Ricci tensor components $R_{00}$ and $R_{11}$ of JNW metric are given by,
\begin{equation}
    R_{00}=0;\quad R_{11}=\frac{b^{2}(1-n^{2})}{2r^{2}(b-r)^{2}}.
\end{equation}
To determine whether Tipler's condition given in Eq. \eqref{Tipler} is satisfied near the singularity, we examine the behavior of $\frac{dr}{d\lambda}$ along radial null geodesics. Following the same procedure as in the JMN-1 case, we construct the Lagrangian describing radial null geodesics in the JNW spacetime,
{\small{\begin{equation}
\mathcal{L}^{\text{null}}_{\text{JNW}}\equiv\frac{1}{2}\Bigg[-\left(1-\frac{b}{r}\right)^{n}\left(\frac{dt}{d\lambda}\right)^{2}+\left(1-\frac{b}{r}\right)^{-n}\left(\frac{dr}{d\lambda}\right)^{2}\Bigg]=0,
\end{equation}}}
and the resulting equation of motion for radial null geodesics becomes 
\begin{equation}
    \frac{d}{d\lambda}\left(\frac{dr}{d\lambda}\right)=0\Rightarrow r(\lambda)= r_{0}+\mathcal{C}_{1}\lambda,
\end{equation}
where $r_{0}$ and $\mathcal{C}_{1}$ are constants. Using the above expression for $r(\lambda)$, we can evaluate the limit defined in Eq.~(\ref{Tipler}), which gives,
\begin{equation} \label{jnwtstr}
\lim_{\lambda\rightarrow0}\lambda^{2}R_{\mu\nu}K^{\mu}K^{\nu}=(1-n^{2})>0.
\end{equation}
This represents that the singularity arising in the JNW spacetime satisfies the Tipler strong curvature condition. In particular, the singularity remains strong within the allowed range $0< n < 1$, implying that any infalling extended object is crushed to zero volume. Moreover, the JMN-1 spacetime singularity exhibits similar behavior for all values of the compactness parameter in the range $M_0 \in (0,1)$. This demonstrates that the singularities in both the JNW and JMN-1 spacetimes are gravitationally strong in the sense of the Tipler criterion. In the next section, we investigate the observational implications of spacelike, null, and timelike singularities in the context of EHT-VLBI imaging.

\section{Connection to EHT Imaging}
\label{sec7}
The remarkable progress in horizon-scale imaging, particularly through the observations of Sgr~A* and M87* by the EHT, has provided unprecedented probes of the spacetime geometry in the strong gravity regime \cite{EventHorizonTelescope:2019dse,EventHorizonTelescope:2022wkp}. While the current data are broadly consistent with the Kerr black hole solution predicted by general relativity \citep{EHTC2022}, they do not yet uniquely rule out all possible alternatives. In particular, certain classes of compact objects can produce observational signatures that closely resemble those expected from Kerr or Schwarzschild spacetimes. These considerations highlight the importance of continued theoretical exploration of compact object solutions both within and beyond general relativity, especially in view of future high-precision observations that may provide more stringent tests of the underlying spacetime geometry \cite{Vagnozzi:2022moj,Ayzenberg:2023hfw}.

It is shown that both JMN-1 and JNW spacetimes can produce shadow structures under specific parameter ranges \cite{Saurabh:2023otl,Shaikh:2018lcc,gyulchev2019,Vagnozzi:2022moj,Bambhaniya2022}. In particular, we have shown here that the JMN-1 geometry exhibits such features for compactness parameter values $2/3 < M_0 < 4/5$, corresponding to the regime in which the singularity is \textit{null} type. Similarly, the JNW spacetime admits shadow formation for $0.5 < n < 1$, where the singularity remains \textit{timelike} in nature but a photon sphere exists. The JMN-1 and JNW can cast shadow images similar to those of a Schwarzschild black hole which have \textit{spacelike} singularity at the center.  This suggests that shadow formation is not uniquely determined by the presence of an event horizon or the causal nature of the singularity, but is closely related to the global behavior of null geodesics and the existence of an upper bound of an effective potential \cite{Bambhaniya:2025iqb}.

We also noticed that recent work by Broderick et al.~\cite{Broderick:2024vjp} argues that many naked singularity models can be excluded observationally because they possess inner turning points for timelike and null geodesics. Such turning points may, in principle, lead to the formation of accretion powered photosphere or shocks within the shadow region, which would contradict EHT observations of Sgr~A* and M87*, where the accretion flow appears coherent up to the photon sphere scale. They conclude that only a restricted class of naked singularities (classified as type P0j \footnote{The P0j-type singularities are defined by a finite angular momentum, where timelike geodesics can reach the singularity. This corresponds to the parameter range in which an unstable photon orbit exists \cite{Broderick:2024vjp}.}), including JMN-1 and JNW spacetimes, remain viable because they allegedly lack these turning points.
However, our analysis shows that this conclusion must be refined. While JMN-1 solutions indeed lack such turning points within the shadow producing regime, the JNW spacetime does admit turning points when $0.5 \leq n < 1$, precisely in the parameter range where shadows exist. The interpretation in~\cite{Broderick:2024vjp}, based on Ref.~\cite{Patil:2011aa,Joshi:2011rlc}, appears to exclude this possibility; however, a closer examination of Ref.~\cite{Patil:2011aa} instead supports our result. Therefore, JNW spacetimes in the shadow producing regime may still allow inner photosphere structures, and their observational viability requires further dedicated modeling rather than immediate exclusion. 

\section{Discussion and Conclusions}
\label{sec8}
In this work, we carried out a detailed causal and dynamical investigation of the JMN-1 and JNW spacetimes by analyzing their light-cone structure, particle turning points, and the corresponding Penrose diagrams. Our analysis shows that the causal nature of the singularities depends sensitively on the underlying parameters of the solutions. In the JMN-1 spacetime, the singularity is null in the range $2/3<M_0<4/5$, whereas it becomes timelike for $0<M_0<2/3$. In contrast, the singularity of the JNW spacetime remains timelike throughout the interval $0<n<1$. 

Despite these distinct causal characters, both geometries are capable of producing shadow images that closely resemble those associated with Schwarzschild black holes within appropriate parameter regimes \cite{Saurabh:2022jjv,gyulchev2019,Vagnozzi:2022moj}. This result has an important conceptual implication. In the Schwarzschild solution, the central singularity is spacelike and hidden behind an event horizon, while in the JMN-1 and JNW spacetimes the singularities are horizonless and may be null or timelike. The existence of a shadow therefore does not depend on the causal type of the singularity, nor on the presence of an event horizon. This conclusion is consistent with recent studies demonstrating that shadow formation is not necessarily tied to the existence of a photon sphere, an event horizon, or even a spacetime singularity \cite{Bambhaniya:2025iqb,Saurabh:2023otl}. 

The strength of the singularities was further examined using Tipler's criterion. Earlier qualitative arguments in the literature have suggested that singularities surrounded by photon spheres may be weak \cite{Paul:2020ufc,Virbhadra:2002ju, Joshi:2011rlc}. The present analysis demonstrates that this expectation does not hold for the spacetimes considered here. Both the JNW and JMN-1 geometries possess strong curvature singularities in the sense of Tipler criterion. In the JNW spacetime, the singularity remains strong throughout the entire parameter range $0<n<1$. A similar result holds for the JMN-1 solution, where the singularity is strong for all $M_0\in(0,0.8)$. These findings show that strong curvature behavior is fully compatible with the formation of shadow images and other observable strong field signatures. As a result, such geometries remain physically relevant as theoretical models of ultra-compact gravitating objects.

The existence of radial turning points for timelike geodesics plays a crucial role in determining the possibility of high-energy particle interactions in these spacetimes. We find that, in the JNW spacetime, such turning points for radially infalling particles exist in the parameter range $0.5 \leq n < 1$, while in the JMN-1 spacetime they occur for $\frac{1}{2} \leq M_0 \leq \frac{2}{3}$. In these regimes, particles originating from infinity with suitable angular momentum can decelerate and reach finite radial turning points within the strong-field region, where they may collide with ingoing particles. These interactions can lead to extremely large center of mass energies, analogous to the BSW mechanism \cite{Patil:2010nt,Patil:2011aa,Patil:2011ya}. Note that such trajectories require a suitable range of angular momentum, and therefore represent a kinematically allowed but not necessarily generic class of particle motion. Therefore, such high-energy collisions may in turn generate shock structures or an effective photosphere, potentially enhancing electromagnetic emission \cite{Broderick:2024vjp}. As a result, observable signatures may arise in EHT shadow imaging, such as excess luminosity or localized bright features near the photon sphere, while the overall shadow structure largely determined by the photon sphere. 

In contrast, for $M_0 > \frac{2}{3}$ in the JMN-1 spacetime, the causal structure undergoes a qualitative change. The strong inward tilting of light-cones suppresses the outward propagation of both photons and massive particles. As a result, photons emitted from the interior fail to reach distant observers, contributing to a shadow appearance, while the absence of outgoing timelike trajectories eliminates radial turning points within the matter region. Without such turning points, particles cannot reverse their motion, and the coexistence of ingoing and outgoing trajectories required for high-energy collisions is no longer possible. Therefore, particle acceleration to large center of mass energies is suppressed within the shadow region, and no shock formation or photosphere like emission is expected for $r \leq 3M$, as determined by the exterior photon sphere.

The present work is formulated within the framework of the test particle approximation, neglecting back-reaction and self-force effects, which may become important at extremely high energies. In realistic astrophysical settings, various dissipative processes such as Coulomb interactions, bremsstrahlung emission, plasma effects, and gravitational redshift can significantly alter the electromagnetic signatures associated with high-energy particle interactions. These mechanisms may modify or suppress features such as excess luminosity or localized bright structures near the photon sphere before the radiation reaches distant observers. A detailed quantitative treatment of these effects is beyond the scope of the present work and is left for future investigation.

Overall, our analysis demonstrates that shadow formation is primarily dictated by the underlying geodesic structure, rather than by the presence of an event horizon or the causal nature of the singularity. Both JMN-1 and JNW spacetimes, characterized by Tipler-strong singularities, provide physically viable strong gravity configurations. Horizonless ultra-compact objects can therefore exhibit observational signatures typically attributed to black holes, while simultaneously allowing for extreme particle dynamics in certain parameter regimes. Incorporating radiative transfer, plasma physics, and realistic observational modeling will be essential in future studies to assess whether such geometries can be distinguished from black holes using next-generation high resolution imaging and timing observations.

\acknowledgments{P. Bambhaniya acknowledge support from the São Paulo State Funding Agency FAPESP (grant 2024/09383-4). A. Bidlan would like to acknowledge the contribution of the COST Action CA23130 (BridgeQG) community. The authors would like to thank A. B. Joshi for providing helpful literature related to Penrose diagrams.}

\nocite{*}
\bibliographystyle{apsrev4-2}
\bibliography{references}

\end{document}